\documentclass[aps,prl,preprint,showpacs,superscriptaddress]{revtex4}

\newcommand{\ts}{\textsuperscript}

\usepackage{graphicx}
\usepackage{amssymb}
\usepackage{amsmath}

\begin{document}

%
%
\title{Bremsstrahlung in $\boldsymbol \alpha$ Decay Reexamined}

\author{H.~Boie}                
\author{H.~Scheit}              
\author{U.~D.~Jentschura}       
\author{F.~K\"ock}                
\author{M.~Lauer}               
\affiliation{Max-Planck-Institut f\"ur Kernphysik, D-69117 Heidelberg, Germany}
\author{A.~I.~Milstein}         
\affiliation{Budker Institute of Nuclear Physics, 630090 Novosibirsk,
  Russia}
\author{I.~S.~Terekhov}         
\affiliation{Budker Institute of Nuclear Physics, 630090 Novosibirsk,
  Russia}
\author{D.~Schwalm}             
\thanks{Present affiliation:  
Department of Particle Physics, Weizmann Institute of Science, Rehovot, Israel}
\affiliation{Max-Planck-Institut f\"ur Kernphysik, D-69117 Heidelberg, Germany}


%
%
\begin{abstract}
  A high-statistics measurement of bremsstrahlung emitted in the
  $\alpha$ decay of \ts{210}Po has been performed, which allows to
  follow the photon spectra up to energies of $\sim{500}$~keV. The
  measured differential emission probability is in good agreement with our
  theoretical results obtained within the quasi classical
  approximation as well as with the exact quantum mechanical
  calculation.  It is shown that due to the small effective electric
  dipole charge of the radiating system a significant interference between the
  electric dipole and quadrupole contributions occurs, which is
  altering substantially the angular correlation between the $\alpha$
  particle and the emitted photon.
\end{abstract}

\maketitle

%
%
The $\alpha$ decay of an atomic nucleus is {\it the} archetypal
quantum mechanical process, and so is the bremsstrahlung accompanied
$\alpha$ decay. It is therefore somewhat surprising that only ten
years ago the first fully quantum mechanical calculation of the latter
process has been performed, using first-order perturbation theory and
the dipole approximation for the photon field ~\cite{PaBe1998}. On the
other hand, it is known that the $\alpha$ decay of a heavy nucleus and
the radiation that accompanies this decay can be treated in the
quasi-classical approximation as well, the applicability of this
approximation being provided by the large value of the Sommerfeld
parameter $\eta$~\cite{DyGo1996,Dy1999}, which amounts to e.g.
$\eta=22$ for the $\alpha$ decay of \ts{210}Po. In all theoretical
approaches the matrix element of bremsstrahlung incorporates
contributions of the classically allowed and classically forbidden
(tunneling) region. The relative contribution of the tunneling region
is not small in general and can be interpreted as bremsstrahlung at
tunneling. However, such an interpretation can only have a restricted
meaning 
as the wavelength of the
photon is much larger than the width of the tunneling region and even
larger than the main classical acceleration region; it is therefore
not possible to identify experimentally the region where the photon
was emitted.  Nevertheless, the issue of tunneling during the emission
process was widely discussed \cite{PaBe1998, DyGo1996, Dy1999,
  TaEtAl1999, Tk1999, BPZ1999, MaOl2003}. The authors used different
theoretical approaches leading to partly conflicting conclusions as to
the contribution of the tunneling process to the bremsstrahlung, but -
more seriously - also with regard to the energy dependent emission
probabilities.

While the theoretical interest in the bremsstrahlung accompanied
$\alpha$ decay was stirred up by an experiment published in
1994~\cite{DAEtAl1994brems}, this and later experimental
attempts~\cite{KaEtAl1997, Eremin00} to observe these
rare decays produced conflicting results and did not reach the
sensitivity to allow for a serious test of the various theoretical
predictions concerning the emission probabilities for
$\gamma$ energies above $E_\gamma \sim{200}$~keV. In the present paper
we report on the first high-statistics measurement of bremsstrahlung
in the $\alpha$ decay of $^{210}$Po, where we have been able to
observe the photon spectra up to $E_\gamma \sim{500}$~keV. Taking into
account the interference between the electric dipole and quadrupole
amplitudes, which we derived within the framework of a refined
quasi-classical approximation~\cite{jen07}, we find good
agreement of our measured $\gamma$ emission probabilities with those
calculated in our quasi-classical approach as well as with the quantum
mechanical prediction of Ref.~\cite{PaBe1998}.

%
%
The main experimental challenge is the very low emission rate for
bremsstrahlung photons. Even with a rather strong $\alpha$ source of
$\sim 100$~kBq the emission rate is only of the order of \textit{one
  per day} in the 300--400~keV energy range, i.e. in only one out of
$10^{10}$ $\alpha$ decays a photon with an energy within that range
will be emitted. Only by measuring the $\alpha$ particles in
coincidence with the bremsstrahlung photons and by identifying the
bremsstrahlung photons by requiring energy balance between the
$\alpha$ energy and the photon one can therefore hope to sufficiently
suppress randoms due to the copious room background.  After a careful
evaluation of possible $\alpha$ emitters, \ts{210}Po, already used in
the work of Ref.~\cite{KaEtAl1997}, was felt to be the most promising
choice for such a measurement. It decays with a halflife of $t_{1/2}=
138$~days mainly to the ground state of the stable daughter nuclide
\ts{206}Pb ($Q_\alpha=5.407$~MeV), with only a small fraction of
$1.22(4) \times 10^{-5}$ proceeding through the first exited
$J^\pi=2^+$ state at an excitation energy of 803~keV~\cite{firestone}.
While no other $\gamma$ rays are emitted by the source, the weak
803~keV branch constitutes a convenient calibration point for the
overall detection efficiency reached in the experiment.

\begin{figure}
  \centerline{\includegraphics[width=8.5cm]{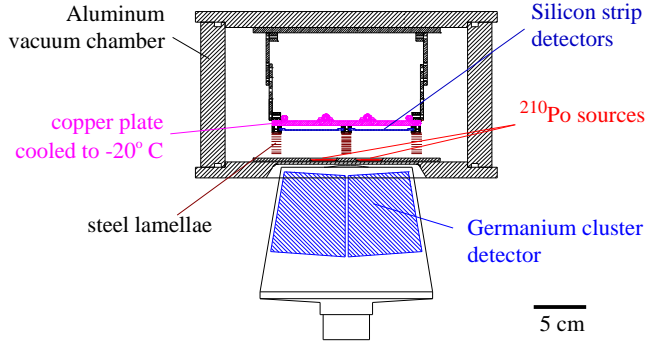}}
  \caption{\label{fig:setup}
    (color online) Cross section of the experimental setup.  }
  \vspace{-0.5cm}
\end{figure}

%
%
The experimental setup used in the present work is shown in
Fig.~\ref{fig:setup}.  Two \ts{210}Po $\alpha$ sources are placed at
the bottom of a common vacuum chamber and are viewed by two segmented
Silicon detectors, each placed about 30 mm above the source to measure
the energy of the $\alpha$ particles.  Directly below the center of
the vacuum chamber an efficient high-purity Germanium triple cluster
detector of the MINIBALL design \cite{eberth:2001} was placed to
record the emitted bremsstrahlung photons.

%
%
The source material was evenly spread on two 0.2~mm thick circular Ni
foils with a diameter of 16~mm, which were mounted on Aluminum disks
of 0.5~mm thickness each. By distributing the source material the $\alpha$
energy loss in the material was minimized; moreover, sputtering of
source material due to the recoil of a nearby $\alpha$ decay was
avoided. The areal uniformity of the activity (100~kBq per source) within
the active area of the source was tested by autoradiography; no intensity
variations could be discerned.

%
The $\alpha$ particles were detected by two $5\times5$ cm$^2$ Silicon
detectors, which were electrically segmented into 16 strips each.
The $\alpha$ particles were incident on the unsegmented side of the
detector in order to avoid events with incomplete charge collection
occurring in the inter-strip region. Both detectors were mounted on a
copper plate cooled to $-20^\circ$~C to improve the energy resolution
and to reduce damage of the Si detectors due to the implanted $\alpha$
particles (about $10^{10}$~/cm$^{2}$ at the end of experiment). The
energy resolution was on the order of 30--35~keV (FWHM) and
deteriorated only for a few strips up to 45~keV at the end of
the production run. The direct path between the left source and the
right Si detector (and vice versa) was mechanically blocked to avoid
large angles of emission and incidence. Moreover, events with two
responding strips were rejected in the off-line analysis, which
considerably improved the low energy tail of the $\alpha$ peaks.
Typical counting rates were $\sim 1.5$~kHz per strip.

The Ge cluster detector is comprised of three large
volume and individually canned Ge crystals, which are housed in a
common cryostat. Each crystal's outer electrode is electrically
segmented into six wedge-shaped segments.  The detector was equipped
with fully digital electronics after the preamplifier, allowing to
record not only the deposited energy but also the signal shapes of the
6 segments and the central core contact of a crystal with a sampling
rate of 40 MHz. The segment hit pattern was used to determine the
(first) interaction point of the photon, and the recorded pulse shapes
were utilized to improve the $\gamma$-$\alpha$ time resolution (FWHM) to
29~ns at $E_\gamma = 100$~keV and 15~ns at $E_\gamma= 500$~keV.  The
$\gamma$ energy was determined by adding the measured core energies of
all three crystals. After carefully shielding the setup with copper
and lead, the counting rate was as low as $\sim 20$~Hz for a threshold
at about 40~keV.

%
%
In view of the spreaded source and the close source-detector geometry,
detailed simulations of the experimental setup were performed to
determine the response function of the cluster detector as well as the
detection efficiency of the setup as a function of $E_\gamma$ and of
the angle $\vartheta$ between the direction of the $\alpha$ particle
and the bremsstrahlung quanta~\cite{hans}. The simulated absolute
$\gamma$ full-energy peak efficiencies were compared to measurements
performed with radioactive sources and to the result deduced from the
803~keV branch of the \ts{210}Po decay; their accuracies were found to
be better than $4 \%$ for $\gamma$ energies above 200~keV but to
deteriorate slightly up to $9 \%$ at 100~keV. The absolute $\gamma$
full-energy peak-efficiency at 803~keV was determined to be
$8.09(26)\%$ for isotropic emission.

%
%
In order to reach the necessary sensitivity the experiment had to be
kept in a stable running condition for many months. The analyzed data
actually correspond to 270 days of data taking with a total of
$4.3\times 10^{11}$ $\alpha$ particles being recorded. At the same
time about $6 \times 10^9$ $\gamma$ rays have been detected out of
which e.g. only about 150 are expected to be due to bremsstrahlung
events in the $\gamma$ energy region above 300~keV.

%
%
\begin{figure}
  \includegraphics[width=8.5cm]{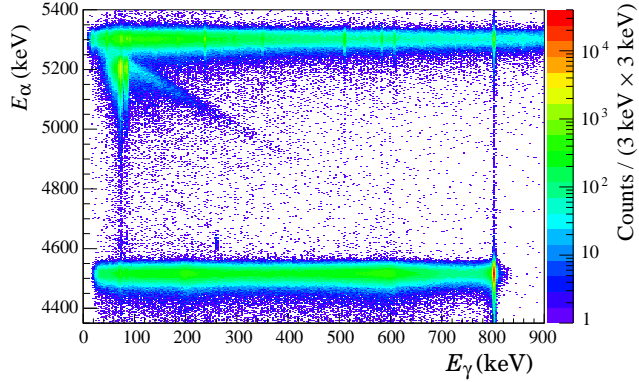}
  \caption{\label{fig:scatter}
    (Color online) Scatter plot showing the energy of the $\alpha$
    particle versus the energy of the $\gamma$ ray, detected
    coincident within a time window of $\pm 50$ ns.  The
    bremsstrahlung events can be clearly discerned along the diagonal
    line starting at $E_\alpha =  E_{\alpha}^0 = 5304$~keV.}  
  \vspace{-0.5cm}
\end{figure}
The $\alpha$-$\gamma$ coincidence matrix displaying the measured
$\alpha$ particle energy versus the $\gamma$-ray energy is shown in
Fig.~\ref{fig:scatter}. The upper horizontal band corresponds to
$\alpha$ particles, which were detected with their full energy of
$E_{\alpha}^0 = 5304$~keV in random coincidence with a background
photon. The lower horizontal band terminating at 803 keV $\gamma$-ray
energy is caused by the response of the Ge detector to the 803 keV
$\gamma$-rays, emitted in coincidence with $\alpha$ particles of
4517~keV leading to the first excited state of \ts{206}Pb. The events
at $\gamma$ energies around between 70~keV and 85~keV for $\alpha$
energies below 5304~keV correspond predominantly to Pb X-rays emitted
after the knockout of an electron by the escaping $\alpha$
particle~\cite{fischbeck}.  Finally, on the diagonal line with
$E_\alpha + (206/210) E_\gamma = \mathrm{const}$ as required by energy
and momentum conservation, bremsstrahlung events can be clearly
observed up to $\gamma$ energies in excess of 400 keV.

\begin{figure}
  \centerline{\includegraphics[angle=-90,width=8.0cm]{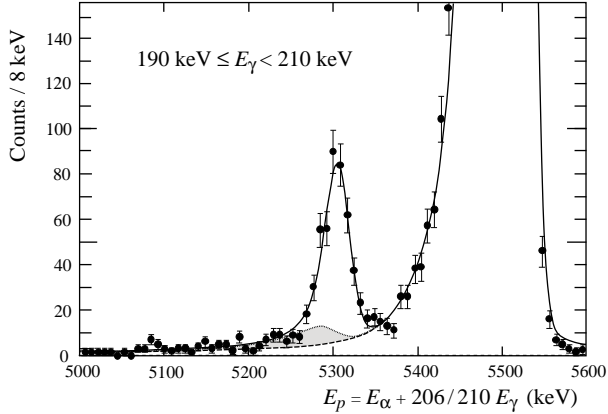}}
  \caption{\label{fig:fit}%
    Projected data and the result of a least-square fit (solid curve)
    for the $\gamma$ energy bin of $190~\rm{keV} \le E_\gamma <210$~keV.
    The peak centered at 5304~keV corresponds to the full-energy peak
    of photons from the bremsstrahlung accompanied $\alpha$ decay,
    while the right structures is due to random $\alpha$-$\gamma$
    coincidences.  }  \vspace{-0.5cm}
\end{figure}

%
%
To determine the differential bremsstrahlung emission probabilities
$dP/dE_\gamma$ coincident $\alpha$-$\gamma$ events with $\gamma$
energies within 20~keV up to 100~keV broad gates were projected along
the diagonal line by plotting them as a function of $E_{p} = E_\alpha
+ (206/210) E_\gamma$. For each $\gamma$ gate the coincidence time
window $\Delta t$ was individually adjusted to include all counts
within $\pm 4.5~\sigma_t(E_\gamma)$ around the centroid of the time
peak with $\sigma_t^2$ denoting its variance. In the projected spectra
the bremsstrahlung events are expected to show up in a sharp peak
around $E_{p} = E_{\alpha}^0 = 5304$~keV independent of the width of
the $\gamma$ energy gate. As an example, the projected energy spectrum
for the 190~keV $\le E_\gamma <$ 210~keV bin is shown in
Fig.~\ref{fig:fit}: Riding on the low energy tail of the random
$\alpha$-$\gamma$ line the bremsstrahlung events are clearly born out.
The solid curve corresponds to the result of a least-squares fit of
the data, where only the intensity and central energy of the
bremsstrahlung peak was varied. The line-shape and intensity of the
random $\alpha$-$\gamma$ peak (dashed line) was determined from the
corresponding projected energy spectrum of the random
$\alpha$-$\gamma$ matrix, scaled by the ratio of the time windows. As
the random matrix has $\sim{10}$ times more statistics, the tail under
the bremstrahlung line could be determined to better than 5$\%$ for
all $\gamma$ energy gates.  The shaded area reflects the Compton
distribution caused by bremsstrahlung events with higher energies. Its
intensity relative to the full energy events and its shape was
determined from the simulation; the accuracy of this contribution is
mainly determined by the intensities of the bremsstrahlungs peaks of the
next higher $\gamma$ gates and is estimated to be $\sim{10}\%$ for the bin
shown in Fig.~\ref{fig:fit}.

From the intensity of the bremsstrahlung line the (solid-angle
integrated) differential emission probability $dP(E_\gamma)/dE_\gamma$
can be determined if the total number of $\alpha$ particles detected
in the Si detectors, $N_{\alpha}^0$, and the probability to detect a
bremsstrahlung photon of energy $E_\gamma$ is known. While
$N_{\alpha}^0$ can be readily determined from the down-scaled single
spectrum recorded during the production run to be $N_{\alpha}^0=
4.311(2)\times 10^{11}$, the detection efficiency requires some extra
considerations as the photons are preferentially detected at backward
angles with respect to the direction of the $\alpha$ particle.  Hence
the $\alpha$-$\gamma$ angular correlation must be taken into account
to extract the differential emission probability.  Usually
bremsstrahlung is assumed to be pure $E1$ radiation as the wavelength
of the emitted radiation is much larger than the dimension of the
radiating system and higher order multipole contributions are
suppressed. However, in the present case the radiating system consists
of the $\alpha$ particle \textit{and} the residual daughter nucleus
\ts{206}Pb, which have rather similar charge to mass ratios such that
the effective dipole charge amounts only to $Z_{E1}^{\mathrm{eff}} =
\mu (z/m-Z/M) = 0.40$, where $z,m$ and $Z,M$ denote the charge and mass
of the $\alpha$ and the daughter nucleus, respectively, and $\mu$ the
reduced mass, while the effective quadrupole charge
$Z_{E2}^{\mathrm{eff}} = \mu^2 (z/m^2 + Z/M^2)=1.95$ is almost a
factor of 5 larger~\cite{jen07}. Even though one does not expect the
$E2$ radiation to contribute sizable to the total, angle integrated
emission probability, the interference between the $E1$ and $E2$
amplitudes might well influence the $\alpha$-$\gamma$ angular
correlation and thus the detection sensitivity of our setup.

\begin{figure}
  \includegraphics[width=8.0cm]{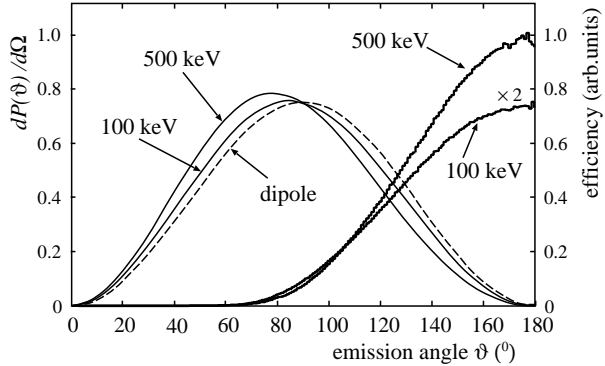}
  \caption{\label{fig:interf} 
    Calculated, normalized $\alpha$-$\gamma$ correlations $dP/d\Omega$
    for bremsstrahlung photons of 100 and 500~keV in comparison to a
    pure dipole correlation. The increasing deviation from the dipole
    characteristic is due to the $E2$ interference. Also shown are the
    emission angle dependent efficiencies (histograms) of our setup
    for photon energies of 100~keV and 500~keV.  }
\end{figure}

We used a refined version of the quasi-classical approach to the
bremsstrahlung emission process~\cite{jen07} to study this question in
more detail, and find indeed a substantial $E2/E1$ interference
contribution to the angular correlation, while the $E2$ contribution
to the angle-integrated differential emission probability is less than
1.5\% at all relevant $\gamma$ energies. Including only the
interference term the angular correlation can be expressed by
$dP(\vartheta)/d\Omega\propto \sin^2\vartheta
(1+2\chi(E_\gamma)\cos\vartheta)$, where $\chi(E_\gamma)$ is
proportional to the ratio of the quadrupole to the dipole matrix
element. In leading order in $1/\eta$ we find that $\chi(E_\gamma)$
approaches zero for $E_\gamma \rightarrow 0$ and increases with
$E_\gamma$ to take e.g. values of +0.09 at 100~keV up to +0.22 at
500~keV.  The influence of the $E2$ amplitude on the angular
correlation is illustrated in Fig.~\ref{fig:interf}, where the
normalized angular correlations of the emitted photons are shown as a
function of $\vartheta$ for bremsstrahlung photons of 100~keV and
500~keV in comparison to a pure dipole emission characteristic
($\chi=0)$.  Multiplied with the acceptance of our setup and
integrated over the solid angle, the $E2$ interference contribution
results in a reduction of the detection efficiency of about 8\% and
24\% at photon energies of 100 and 500 keV, respectively, when
compared to the efficiency expected for a pure dipole emission. From
an estimate of the next to leading order contribution to $\chi$ we can
estimate the uncertainty to the detection efficiency caused by using
only the leading order term in the angular correlation to be less than
3\%.

%
%

\begin{figure}
  \includegraphics[angle=-90,width=8.5cm]{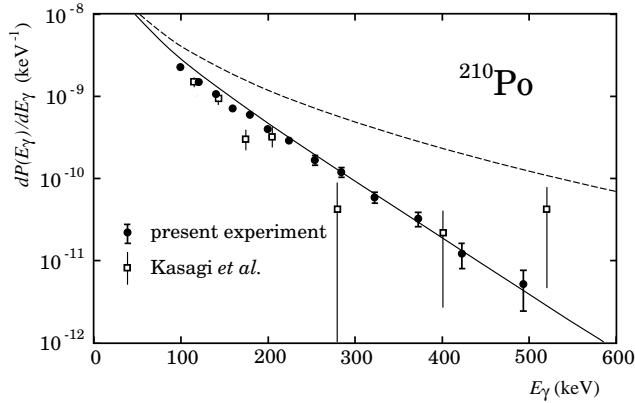}
  \caption{\label{fig:result}
    Differential Bremsstrahlung emission probability
    $dP(E_\gamma)/dE_\gamma$ for the decay of \ts{210}Po (solid
    points: present work, open symbols: Kasagi {\it et
      al.}~\cite{KaEtAl1997}).  The solid curve reflects the
    calculation of Papenbrock and Bertsch~\cite{PaBe1998} as well as
    the result of our quasi-classical calculation~\cite{jen07}. The
    result of a classical Coulomb acceleration calculation is shown by
    the dashed line. }  \vspace{-0.2cm}
\end{figure}

The resulting solid-angle integrated and efficiency corrected
differential emission probabilities $dP(E_\gamma)/dE_\gamma$ are displayed
in Fig.~\ref{fig:result} by the solid points. The $1\sigma$ errors
shown comprise the statistical and systematic uncertainties and are
smaller than the point size for $\gamma$ energies below 250~keV; they
amount to e.g. $\sigma_{sta} = 2 \%$ and $\sigma_{sys} = 8 \%$ at
$<E_\gamma> = 139$~keV, and $\sigma_{sta} = 19 \%$ and $\sigma_{sys} = 5
\%$ at $<E_\gamma> = 373$~keV.
Also shown are the earlier results obtained by Kasagi {\it et
  al.}~\cite{KaEtAl1997}.  Note, that external bremsstrahlung
contributions, which stem from the slowing down of the $\alpha$
particles in the Si detector material, are several orders of magnitude
smaller than the measured probabilities.

%
%
%

In Fig.~\ref{fig:result} our data are also compared with the
predictions of Papenbrock and Bertsch~\cite{PaBe1998} and of our
quasi-classical approach~\cite{jen07}; the two theoretical approaches
actually agree with each other to better than 2\% as shown
in~\cite{jen07} and are thus indistinguishable on the scale of
Fig.~\ref{fig:result}.  Overall, very good agreement between theory
and experiment is observed, however, small deviations of up to 20\%
are encountered at energies below 200~keV, which are also supported by
the data of Kasagi {\it et al.}~\cite{KaEtAl1997}. It remains to be
seen if these deviations can be traced back to the use of the
potential model so far employed in all theoretical calculations to
describe the interaction of the $\alpha$ particle and the daughter
nucleus at distances of the order of the nuclear radius. The present
high precision data clearly demonstrates the failure of a classical
Coulomb acceleration calculation (see e.g.~\cite{KaEtAl1997,hans}) to
describe the bremsstrahlung emission in $\alpha$ decay, and rules out
theoretical suggestions put forward by the authors of
Refs.~\cite{KaEtAl1997,MaOl2003,BPZ1999}.

U.D.J. acknowledges support from the Deutsche Forschungsgemeinschaft
(Heisenberg program) and D.S. support by a Joseph Meyerhoff Visiting
Professorship granted by the Weizmann Institute of Science.  A.I.M. and
I.S.T.  gratefully acknowledge the Max-Planck-Institute for Nuclear
Physics, Heidelberg, for warm hospitality and support. The work was
also supported by RFBR Grant No.~03-02-16510.

%
%

\end{document}